\documentclass[aps,prb,twocolumn,showpacs]{revtex4}
\usepackage{graphicx}

\begin{document}

\title{High-$T_{c}$ superconductivity induced by doping rare earth elements into CaFeAsF}

\author{Peng Cheng, Bing Shen, Gang Mu, Xiyu Zhu, Fei Han, Bin Zeng and Hai-Hu Wen}\email{hhwen@aphy.iphy.ac.cn }

\affiliation{National Laboratory for Superconductivity, Institute of
Physics and Beijing National Laboratory for Condensed Matter
Physics, Chinese Academy of Sciences, P. O. Box 603, Beijing 100190,
China}
\date{\today}

\begin{abstract}
We have successfully synthesized the fluoride-arsenide compounds
Ca$_{1-x}$RE$_x$FeAsF (RE=Nd, Pr; x=0, 0.6). The x-ray powder
diffraction confirmed that the main phases of our samples are
Ca$_{1-x}$RE$_x$FeAsF with the ZrCuSiAs structure. By measuring
resistivity, superconductivity was observed at 57.4 K in Nd-doped
and 52.8 K in Pr-doped samples with x=0.6. Bulk superconductivity
was also proved by the DC magnetization measurements in both
samples. Hall effect measurements revealed hole-like charge carriers
in the parent compound CaFeAsF with a clear resistivity anomaly
below 118 K, while the Hall coefficient $R_H$ in the normal state is
negative for the superconducting samples Ca$_{0.4}$Nd$_{0.6}$FeAsF
and Ca$_{0.4}$Pr$_{0.6}$FeAsF. This indicates that the rare earth
element doping introduces electrons into CaFeAsF which induces the
high temperature superconductivity.
\end{abstract} \pacs{74.70.Dd, 74.25.Fy, 75.30.Fv, 74.10.+v}
\maketitle

Since the discovery of superconductivity in the quaternary compound
LaFeAsO$_{1-x}$F$_x$ with $T_c$ = 26 K,\cite{Kamihara2008} the quest
for new high-$T_c$ superconductors in this FeAs-based family has
never ceased. Especially when replacing La with other rare earth
elements, a group of superconductors with critical temperature well
exceeding 50 K were fabricated,\cite{Pr52K,RenZA55K,WangC} which
excites the whole physical society. Besides these doped REFeAsO (RE
= rare earth elements) superconductors with ZrCuSiAs-type structure
(abbreviated as the FeAs-1111 phase), iron-based superconductors
with different structures were also synthesized, such as (Ba,
Sr)$_{1-x}$K$_x$Fe$_2$As$_2$ (ThCr$_2$Si$_2$-type,
FeAs-122),\cite{BaKparent,Rotter,CWCh} Li$_x$FeAs (PbFCl-type,
FeAs-111),\cite{LiFeAs,LiFeAsChu,LiFeAsUK} FeSe (PbO-type,
FeAs-11).\cite{FeSeMK} Among them the superconductors with the
FeAs-1111 phase seems to have the highest superconducting transition
temperature. The highest T$_c$ reported so far is about 56 K in
Gd$_{1-x}$Th$_x$FeAsO.\cite{WangC} Very recently a new series of
FeAs-based compounds were successfully synthesized, namely
fluoride-arsenides AEFeAsF (AE = divalent metals: Ca, Sr, Eu) with
the ZrCuSiAs
structure.\cite{CaFhosono,SrFJohrendt,SrFxiyuzhu,SrFfeihan} This new
compound is an analogue of REFeAsO, where the (REO)$^+$ layer is
replaced by (AEF)$^+$ layer. For SrFeAsF, superconductivity could
emerge by either doping cobalt directly into the
FeAs-layer\cite{SrCohosono} or partial replacement of Sr with rare
earth elements (Lanthanum\cite{SrFxiyuzhu},
Samarium\cite{CaSmxhchen}). While for CaFeAsF, superconductivity was
also realized by doping 3d transition metals (Co and Ni) into the
iron sites in the FeAs-layer with the highest T$_c$ of about 22 K in
the case of Co doping.\cite{CaFhosono,3dhosono} This relative low
value of $T_c$ could be due to the damage to the FeAs planes when
substituting Fe with Co. Thus it is interesting to know whether the
rare earth elements doping could get superconductors with higher
critical temperatures since it is the case in La-doped
SrFeAsF\cite{SrFxiyuzhu} where superconductivity at about 32 K was
observed. In this paper we report the discovery of high-temperature
superconductivity in Ca$_{0.4}$RE$_{0.6}$FeAsF with T$_c$(onset) =
57.4 K by doping Nd and 52.8 K by doping Pr. The transition
temperature at 57.4 K marks the highest record so far in the
FeAs-based superconductors.

\begin{figure}
\includegraphics[width=9cm]{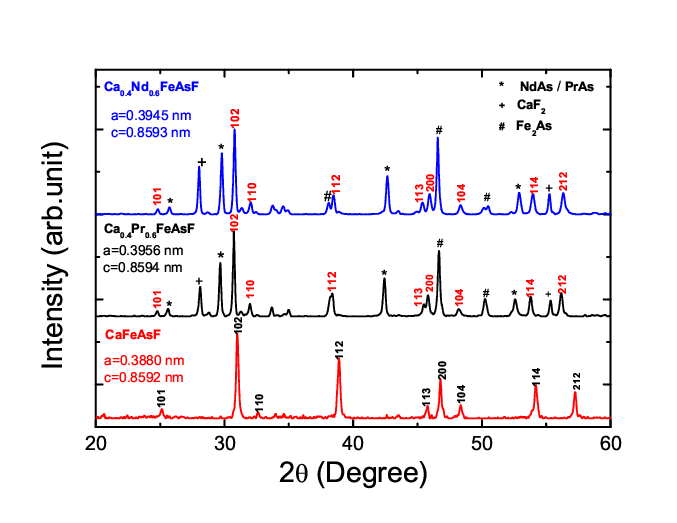}
\caption{(Color online) X-ray diffraction patterns for
Ca$_{1-x}$RE$_x$FeAsF (RE=Nd, Pr; x=0, 0.6) samples. All the main
peaks can be indexed to the tetragonal ZrCuSiAs-type structure. For
the doped samples, the peaks from the impurities are precisely
indexed to the phases NdAs (PrAs), CaF$_2$ and Fe$_2$As. }
\label{fig1}
\end{figure}

The Ca$_{1-x}$RE$_x$FeAsF samples were synthesized by solid-state
reaction method. Firstly, CaAs, NdAs and PrAs were prepared by
heating Calcium pieces (purity 99.99\%), Pr pieces (purity 99.99\%)
and Nd pieces (purity 99.99\%) with As powder (purity 99.99\%)
respectively at 700 $^o$C for 10 hours. Then stoichiometric CaAs,
REAs(RE = Nd or Pr), FeF$_2$ (purity 99\%) and iron powder (purity
99.99\%) were mixed as the nominal composition Ca$_{1-x}$RE$_x$FeAsF
(x=0, 0.6), grounded and pressed into a pellet. All the processes
were carried out in a glove box with argon atmosphere (both H$_2$O
and O$_2$ are limited below 0.1 ppm). Finally the product was sealed
in a quartz tube with 0.4 bar of high purity Ar gas. It was then
slowly heated up to and stayed at 900 $^o$C for 20 hours and
followed by a treatment at 1050 $^o$C for 20 hours.

\begin{figure}
\includegraphics[width=9cm]{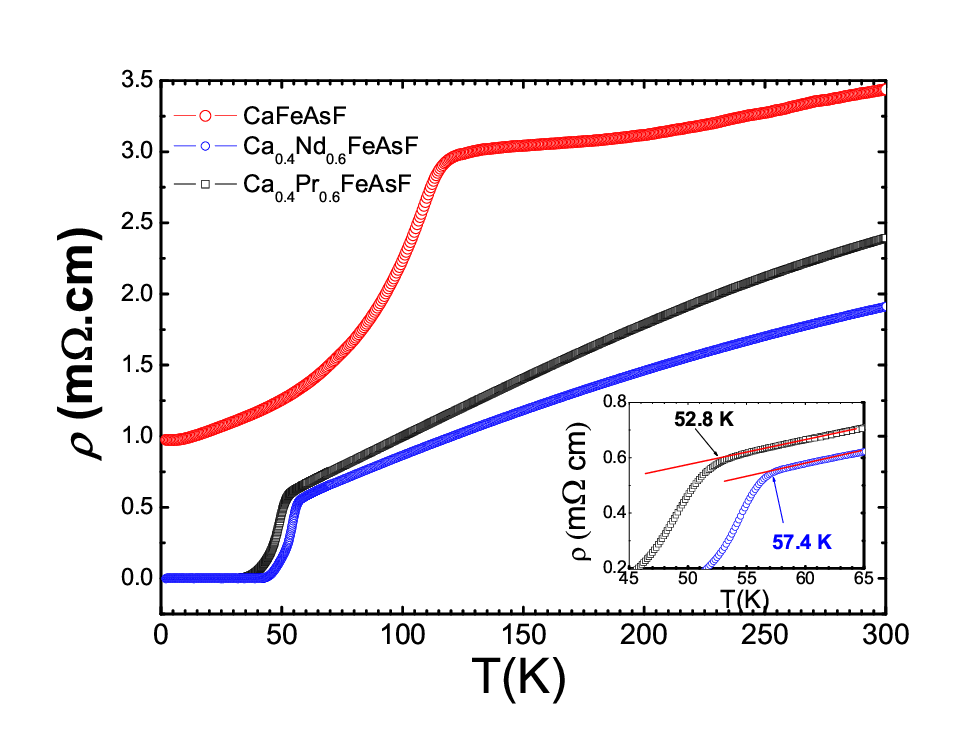}
\caption{(Color online) Temperature dependence of resistivity for
the CaFeAsF and Ca$_{0.4}$RE$_{0.6}$FeAsF(RE=Nd, Pr). The inset
shows an enlarged view in the region of the superconducting
transition.} \label{fig2}
\end{figure}

The x-ray diffraction measurement was performed at room temperature
using an MXP18A-HF-type diffractometer with Cu-K$_{\alpha}$
radiation from 10$^\circ$ to 60$^\circ$ with a step of 0.01$^\circ$.
The analysis of x-ray powder diffraction data was done by using the
software POWDER-X\cite{DongC}. The DC magnetization measurements
were carried out on a Quantum Design superconducting quantum
interference device (SQUID) magnetometer. The resistance and
Hall-effect data were collected using a six-probe technique on a
Quantum Design instrument physical property measurement system
(PPMS) with magnetic fields up to 9 T.

In Fig. 1 we show the x-ray diffraction (XRD) patterns for the
samples Ca$_{1-x}$RE$_x$FeAsF (x=0, 0.6). One can see that the
parent phase CaFeAsF is rather pure where all diffraction peaks can
be indexed by the tetragonal structure with a=0.3880 nm and c=0.8592
nm. When Nd or Pr was introduced into this system some secondary
phases appear, which is similar to the case in Samarium-doped
SrFeAsF.\cite{CaSmxhchen} This is understandable since CaF$_2$ is a
very stable compound which can be easily formed during the reaction.
The left off-stoichiometric compositions will lead to the formation
of Ca$_{1-x}$RE$_x$FeAsF, Fe$_2$As and NdAs or PrAs. Therefore it
would be helpful for forming the superconducting phase by blocking
the formation of the CaF$_2$ phase and thus the Fe$_2$As and REAs
(RE=Nd and Pr). From Fig.1 one can see that all the main peaks of
RE-doped samples could be well indexed to a ZrCuSiAs-type structure
with a=0.3945 nm and c=0.8593 nm for Ca$_{0.4}$Nd$_{0.6}$FeAsF and
a=0.3956 nm and c=0.8594 nm for Ca$_{0.4}$Pr$_{0.6}$FeAsF, while all
the other peaks were indexed precisely to the standard XRD patterns
of NdAs(PrAs), CaF$_2$ and Fe$_2$As. We all know that the radius of
Pr$^{3+}$ is around 1.01 $\AA$ and Nd$^{3+}$ is 1.0 $\AA$ which are
both a little larger than that of Ca$^{2+}$ (0.99 $\AA$), so the
lattice parameters may expand when rare earth elements were doped
into the lattice. Compared to CaFeAsF, the c-axis lattice constants
of RE-doped sample does not increase obviously while the a-axis
lattice constant suffers a major change (increases about 1.6\% for
Nd-doped samples). The similar phenomena was also observed in
La-doped SrFeAsF\cite{SrFxiyuzhu} and Sm-doped
SrFeAsF.\cite{CaSmxhchen} One may argue that the superconductivity
here could be induced by the F-doped REFeAsO phase which might
happen if some amount of oxygen leaked into the samples. This
possibility can be however ruled out by the following arguments.
Firstly, the weighing, mixing and pressing procedures were performed
strictly in the glove box which limits the oxygen down to a
undoubted low level. The dense pellet was sealed into a quartz tube
quickly afterwards. There is no chance for much oxygen going into
the sample. Secondly, we can also get a support from the detailed
analysis on the x-ray diffraction data. Taking Pr-doping as an
example, the typical lattice constants of F-doped PrFeAsO with T$_c$
= 52 K were: a=0.3967 nm and c=0.8561 nm,\cite{Pr52K} which could
not match any main peaks in our x-ray diffraction data. It is the
same case if we compare the lattice constants of the F-doped NdFeAsO
and the data from our sample Ca$_{0.4}$Nd$_{0.6}$FeAsF. Since all
the other impurity peaks could be accurately indexed to the standard
XRD patterns of PrAs, CaF$_2$ and Fe$_2$As (as shown in Fig.1), no
traces of F-doped (Nd or Pr)FeAsO could be detected in our samples.

Fig.2 shows the temperature dependence of resistivity for samples
Ca$_{1-x}$RE$_x$FeAsF (RE=Nd, Pr; x=0, 0.6). A clear resistivity
anomaly was observed at about $T_{an}$=118 K for the parent sample
which could be attributed to the structural phase transition or an
anti-ferromagnetic order. For the superconducting samples this
anomaly could not be seen and a superconducting transition appears
at 57.4 K (onset) for Nd-doped sample and 52.8 K (onset) for the
Pr-doped one. The normal state resistivity of the superconducting
samples shows a metallic behavior in wide temperature region which
is typical in these optimally doped FeAs-1111
superconductors.\cite{SrFxiyuzhu}

\begin{figure}
\includegraphics[width=9cm]{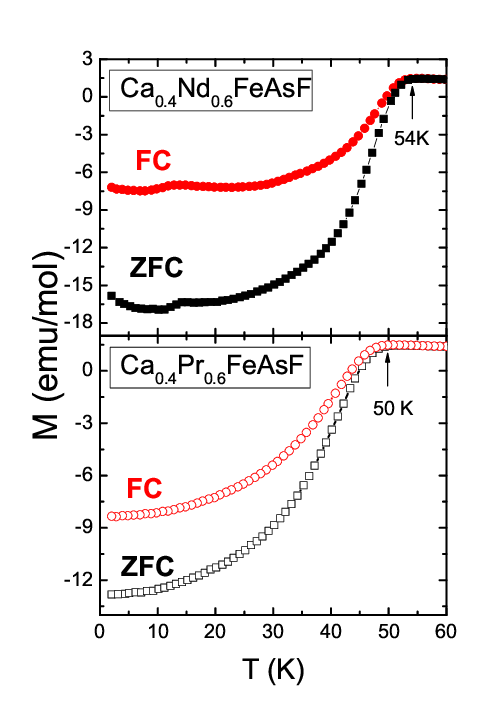}
\caption{(Color online)  Temperature dependence of DC magnetization
for samples Ca$_{0.4}$Nd$_{0.6}$FeAsF and Ca$_{0.4}$Pr$_{0.6}$FeAsF.
A DC field of 50 Oe was applied in the measurements with the zero
field cooling and field cooling modes.} \label{fig3}
\end{figure}

The bulk superconductivity of Ca$_{0.4}$RE$_{0.6}$FeAsF samples were
confirmed by DC magnetization measurement under a magnetic field of
50 Oe in zero field cooling and field cooling processes. As shown in
Fig.3, the onset point of the diamagnetic transition locates at
about 54 K for Nd-doped sample and 50 K for Pr-doped one. Both
samples exhibit a positive background in the normal state which may
come from the impurity phase Fe$_2$As. It is interesting to note
that a small kink appears at about 14 K in the curve of M vs. T in
all superconducting samples doped with Nd. But it has never been
observed in the samples with Pr doping. Therefore we attribute this
effect to the possible existence of the AF ordering of the Nd$^{3+}$
ions because they have the magnetic moments. While for Pr$^{3+}$
ions no magnetic moment is anticipated.

\begin{figure}
\includegraphics[width=9cm]{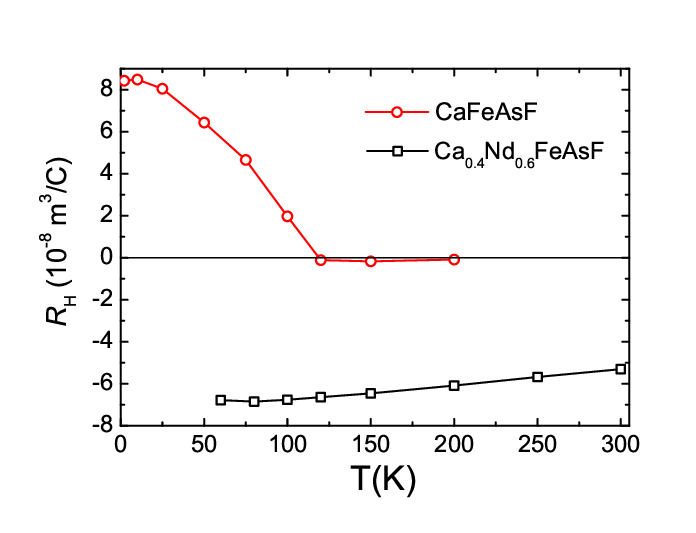}
\caption{(Color online)  Temperature dependence of Hall coefficients
$R_{H}$ for samples CaFeAsF and Ca$_{0.4}$Nd$_{0.6}$FeAsF. The
measurement was done through sweeping magnetic field at fixed
temperatures.} \label{fig3}
\end{figure}

To investigate the evolution of charge carriers we conducted Hall
effect measurements on our samples. For the parent compound, as
shown in Fig.4, the sign of Hall coefficient changes at $T_{an}$=118
K. When the temperature is above $T_{an}$, $R_{H}$ is small but
negative and does not change much with temperature. While below 118
K, $R_{H}$ becomes positive and increases quickly when temperature
decreases. This phenomena could be attributed to the emergence of a
spin-density-wave order in lower temperature region. Band structure
calculations showed that the Fermi surfaces of CaFeAsF is similar to
that in LaFeAsO with two electron pockets in $M$ point and two hole
pockets at $\Gamma$ point\cite{Russia}, the SDW order removes the
density of states on some Fermi pockets and may leave one of the
hole pockets partially of fully ungapped thus causes the low
temperature behavior of $R_{H}$\cite{SrFfeihan}. Through partial
substitution of Ca$^{2+}$ with Nd$^{3+}$ or Pr$^{3+}$, electrons
were introduced into this system. As shown in Fig.4, the Hall
coefficient $R_{H}$ was found to be negative in
Ca$_{0.4}$RE$_{0.6}$FeAsF, which confirmed that the electron-type
charge carriers were doped into the system by substituting Ca
partially with the tri-valent rare earth elements Nd and Pr. It
should be mentioned that superconductivity was not observed in
samples with low doping ($x\leq0.2$), which may be understood that
the parent phase has a hole pocket partially or fully un-gapped,
thus one needs to doped more electrons into the system in order to
carry out the superconductivity. If this is the case, it would be
optimistic to induce superconductivity by a small amount of hole
doping. This is just underway. We have so far made tens of samples
which show the superconductivity beyond 50 K in the fluoride-based
systems when doping about 40-60\% Nd or Pr for Ca. Further efforts
are worthwhile in order to get rid of the impurity phase, especially
to block the formation of CaF$_2$ during the reaction. It remains to
be found out how high the superconducting transition temperature
would go if we substitute the Ca with other rare earth elements.

In summary, high-temperature superconductivity was observed in
compounds Ca$_{0.4}$RE$_{0.6}$FeAsF (RE=Pr, Nd). The onset
superconduting transition temperature is about 57.4 K as determined
from the resistivity data in Ca$_{0.4}$Nd$_{0.6}$FeAsF. DC
magnetizations measurement confirmed the bulk superconductivity of
our samples. The measurements of Hall coefficients $R_{H}$ reveal
the evolution of charge carriers from hole-type in the parent phase
CaFeAsF to electron-type in Ca$_{0.4}$Nd$_{0.6}$FeAsF. Our results
show that high temperature superconductivity could emerge through
rare earth element doping into (CaF)$^+$ layer in this CaFeAsF
system.

We are grateful to Dr. H. Eisaki and A. Iyo for useful discussions
during the visit of WHH in the lab of AIST. We also acknowledge the
help by L. H. Yang and H. Chen in measuring the x-ray diffraction
data. This work is supported by the Natural Science Foundation of
China, the Ministry of Science and Technology of China (973 project:
2006CB01000, 2006CB921802), the Knowledge Innovation Project of
Chinese Academy of Sciences.

\end{document}